\documentclass[%
    preprint,
    amsmath,
    amssymb,
    aps,
    prm,
    showkeys,
    superscriptaddress,
]{revtex4-2}
\usepackage{graphicx}
\usepackage{dcolumn}
\usepackage{bm} 
\usepackage{hyperref} 

\usepackage{textgreek} 
\usepackage{chemformula} 
\usepackage{siunitx} 
\usepackage{multirow} 
\usepackage{rotating} 

\usepackage{xcolor} 
\usepackage{lipsum} 

\DeclareFontShape{OT1}{cmss}{m}{it}{<->ssub*cmss/m/sl}{} 
\usepackage[cm]{sfmath} 

\DeclareSIUnit\angstrom{\text{Å}}   
\DeclareSIUnit\atm{\text{atm}}      
\DeclareSIUnit\amu{\text{u}}        

\newcommand{\hbasym}{\Delta r_{\mathrm{HB}}}
\newcommand{\epsA}{\epsilon_{\ch{A2}}}
\newcommand{\epsMX}{\epsilon_{\ch{MX4}}}
\newcommand{\epsMXpar}{\epsMX^{\parallel}}
\newcommand{\epsMXperp}{\epsMX^{\perp}}
\newcommand{\sigmatwo}{\sigma^{2}}
\newcommand{\deltad}{\Delta d}
\newcommand{\degreechi}{\chi^{y}}
\newcommand{\phiS}{\phi^{S}}
\newcommand{\phiR}{\phi^{R}}

\newcommand{\vectorchirality}{\bm{\epsilon}}
\newcommand{\chirality}{\epsilon}

\usepackage{mathtools} 

\DeclarePairedDelimiterX\braket[2]{\langle}{\rangle}{#1\,\delimsize\vert\,\mathopen{}#2}

\newcommand{\phononcircpolgeneral}{s^{\alpha}_{\mathbf{q},\sigma}}

\newcommand{\triplescrew}{$P2_{1}2_{1}2_{1}$}

\newcommand{\mbapb}{\ch{MBA2PbI4}} 
\newcommand{\smbapb}{(\textit{S}-\ch{MBA)2PbI4}} 
\newcommand{\rmbapb}{(\textit{R}-\ch{MBA)2PbI4}} 
\newcommand{\racmbapb}{(\textit{rac}-\ch{MBA)2PbI4}} 

\newcommand{\mbasn}{\ch{MBA2SnI4}} 
\newcommand{\smbasn}{(\textit{S}-\ch{MBA)2SnI4}} 
\newcommand{\rmbasn}{(\textit{R}-\ch{MBA)2SnI4}} 
\newcommand{\racmbasn}{(\textit{rac}-\ch{MBA)2SnI4}} 

\newcommand{\mbamix}{\ch{MBA2Sn_{x}Pb_{1-x}I4}} 
\newcommand{\smbamix}{(\textit{S}-\ch{MBA)2Sn_{x}Pb_{1-x}I4}} 

\newcommand{\smbamixhalf}{(\textit{S}-\ch{MBA)2Sn_{0.5}Pb_{0.5}I4}} 

\begin{document}

\title{Impact of Metal Cation on Chiral Properties of 2D Halide Perovskites}

\author{Mike Pols}
\email{m.c.w.m.pols@tue.nl}
\affiliation{%
    Materials Simulation \& Modelling, Department of Applied Physics and Science Education, Eindhoven University of Technology, 5600 MB, Eindhoven, The Netherlands
}%

\author{Helena Boom}
\affiliation{%
    Materials Simulation \& Modelling, Department of Applied Physics and Science Education, Eindhoven University of Technology, 5600 MB, Eindhoven, The Netherlands
}%

\author{Geert Brocks}
\affiliation{%
    Materials Simulation \& Modelling, Department of Applied Physics  and Science Education, Eindhoven University of Technology, 5600 MB, Eindhoven, The Netherlands
}%
\affiliation{%
    Computational Chemical Physics, Faculty of Science and Technology and MESA+ Institute for Nanotechnology, University of Twente, 7500 AE, Enschede, The Netherlands
}%

\author{Sof\'{i}a Calero}
\affiliation{%
    Materials Simulation \& Modelling, Department of Applied Physics and Science Education, Eindhoven University of Technology, 5600 MB, Eindhoven, The Netherlands
}%

\author{Shuxia Tao}
\email{s.x.tao@tue.nl}
\affiliation{%
    Materials Simulation \& Modelling, Department of Applied Physics and Science Education, Eindhoven University of Technology, 5600 MB, Eindhoven, The Netherlands
}%

\begin{abstract}

Chiral two-dimensional (2D) halide perovskites are formed by embedding chiral organic cations in a perovskite crystal structure. The chirality arises from distortions of the 2D metal halide layers induced by the packing of these organic cations. \ch{Sn}-based octahedra spontaneously distort, but it remains unclear whether this intrinsic structural instability enhances the chirality. We investigate the effect of the metal cation on structural and phonon chirality in \mbamix{} (x = 0, 1/2, and 1). Incorporating \ch{Sn} does distort the metal halide octehedra, yet it only has a minor impact on the structural chirality. In contrast, the phonons in \mbasn{} are substantially more chiral than in \mbapb{}, especially the in-plane acoustic modes. However, this enhanced phonon chirality does not lead to a generation of a larger angular momentum under a temperature gradient, because the contributions of different chiral phonons tend to compensate one another.

\end{abstract}


\maketitle

\section*{Introduction}

Compositional engineering is a core advantage of mixed organic–inorganic halide perovskites, offering tunable control over structural, electronic, and optical properties~\cite{jeonCompositionalEngineeringPerovskite2015, bushCompositionalEngineeringEfficient2018, taoAbsoluteEnergyLevel2019, manninoTemperatureDependentOpticalBand2020, iqbalCompositionDictatesOctahedral2024}. For example, substituting iodide with bromide or chloride allows systematic modification of the band gaps and band offsets, exciton binding energies, and carrier dynamics~\cite{colellaMAPbI3xClxMixedHalide2013, bokdamRolePolarPhonons2016, nasstromDependencePhaseTransitions2020}. Such versatility enables tailored performance for diverse optoelectronic applications, including solar cells~\cite{greenEmergencePerovskiteSolar2014}, light-emitting diodes (LEDs)~\cite{liuMetalHalidePerovskites2021}, and photodetectors~\cite{wangLowDimensionalMetalHalide2021}.

The incorporation of larger organic cations into the crystal structure breaks up the three-dimensional perovskite lattice, resulting in so-called two-dimensional (2D) perovskites~\cite{maoTwoDimensionalHybridHalide2019, li2DHalidePerovskite2021}. In these materials, 2D metal halide layers alternate with layers of organic cations. The organic cations offer a way to introduce new functionalities. In particular, chiral organic cations can induce structural distortions~\cite{janaOrganictoinorganicStructuralChirality2020, sonUnravelingChiralityTransfer2023, polsTemperatureDependentChiralityHalide2024}, that enable chiral properties such as circular dichroism (CD)~\cite{ahnNewClassChiral2017, dangBulkChiralHalide2020}, chirality-induced spin splitting (CISS)~\cite{luSpindependentChargeTransport2019, kimChiralinducedSpinSelectivity2021}, or chiral heat transport~\cite{kimChiralphononactivatedSpinSeebeck2023, polsChiralPhonons2D2025}.

To achieve such chiral distortions, the metal halide octahedra are strained through the packing of the chiral cations. In the absence of such strain, the metal halide octahedra have a high symmetry~\cite{fuStereochemicalExpressionNs22021}, which prohibits chirality~\cite{janaOrganictoinorganicStructuralChirality2020}. Tin halide perovskites commonly display a lower symmetry, as \ch{Sn^{2+}} ions have a prominent stereochemically active lone pair, which can break the octahedral symmetry of the halide positions surrounding the tin ion~\cite{swainsonOrientationalOrderingTilting2010, swiftLonePairStereochemistryInduces2023}. Even when such distortions are not present in the static structure, a dynamic off-centering of the \ch{Sn} cations within the metal halide octahedra can take place, an effect called \textit{emphanisis}~\cite{fabiniDynamicStereochemicalActivity2016, lauritaChemicalTuningDynamic2017}.

In contrast, such local distortions are generally absent in lead-based analogues, as the lone pair on \ch{Pb^{2+}} ions is significantly less stereochemically active~\cite{fuStereochemicalExpressionNs22021}. While research on metal cation off-centering has mostly focused on three-dimensional (3D) perovskites, there is increasing evidence that similar symmetry breaking also occurs in 2D perovskites~\cite{mariaEvidenceLonePair2024}. In particular, recent work on chiral 2D perovskites, \smbamix{}, has shown that structural distortions strongly affect the local geometry, where \smbasn{} exhibits a markedly greater symmetry breaking compared to \smbapb{}~\cite{luHighlyDistortedChiral2020}.

Given the tendency of \ch{Sn} halide octahedra to distort, combined with the effect it has on the optoelectronic properties~\cite{fortinoAtomisticModelingMetal2023}, it is sensible to ask whether this intrinsic structural instability can aid in boosting chirality. To answer this question, we systematically investigate the static and dynamic chirality in chiral 2D perovskites with varying metal cation compositions (i.e. \mbamix{}). First, we quantify static octahedral distortions and structural chirality, using descriptors developed in earlier work~\cite{polsTemperatureDependentChiralityHalide2024}. Next, we explore dynamic chirality, specifically its temperature dependence, by means of molecular dynamics (MD) simulations with on-the-fly machine-learning force fields (MLFFs), trained on data obtained from density functional theory (DFT) calculations. Finally, we analyze the lattice vibrations to resolve phonon chirality and assess the phonon angular momentum under a thermal gradient.

Our results give rise to several conclusions. Replacing \ch{Pb} by \ch{Sn} indeed substantially distorts the metal halide octahedra, but the effect of this substitution on the structural chirality is much smaller. The effect of temperature on the chirality is similar for \ch{Sn}- and \ch{Pb}-based perovskites. The low-energy phonon modes in \smbasn{} exhibit substantially greater chirality than those in \smbapb{}, confirming that \ch{Sn} substitution enhances dynamic chirality. However, the angular momentum generated during heat transport is larger in the \ch{Pb}-based perovskite because, in \smbasn{}, chiral acoustic and optical phonon contributions largely cancel out, while in \smbapb{} the acoustic modes dominate.

\section*{Methods}

\subsection*{Density functional theory}

Density functional theory (DFT) calculations are performed using the Vienna Ab-initio Simulation Package (\texttt{VASP})~\cite{kresseInitioMoleculardynamicsSimulation1994, kresseEfficiencyAbinitioTotal1996, kresseEfficientIterativeSchemes1996}. Valence electrons for \ch{H} (1s\textsuperscript{1}), \ch{C} (2s\textsuperscript{2}2p\textsuperscript{2}), \ch{N} (2s\textsuperscript{2}2p\textsuperscript{3}), \ch{Sn} (5s\textsuperscript{2}5p\textsuperscript{2}), \ch{I} (5s\textsuperscript{2}5p\textsuperscript{5}), and \ch{Pb} (6s\textsuperscript{2}6p\textsuperscript{2}) are treated using projector-augmented wave (PAW) pseudopotentials~\cite{kresseUltrasoftPseudopotentialsProjector1999}. The strongly constrained and appropriately normed (SCAN) functional is used to model electron-electron exchange-correlation interactions~\cite{sunStronglyConstrainedAppropriately2015}. A plane-wave basis set with a cutoff energy of \SI{500}{\eV} is employed for all calculations, and a $\Gamma$-centered $k$-mesh is used to sample the reciprocal space~\cite{monkhorstSpecialPointsBrillouinzone1976}. Structural optimizations are carried out by relaxing all atomic positions, cell shape, and cell volume until the energy and force convergence criteria of 10\textsuperscript{-5} \SI{}{\eV} and 10\textsuperscript{-2} \SI{}{\eV\per\angstrom}, respectively, are achieved.

\subsection*{Machine-learning force fields}

Kernel-based machine-learning force fields (MLFFs) are trained using total energies, forces, and stresses obtained from DFT calculations. The training set structures are generated through an on-the-fly active learning algorithm implemented in \texttt{VASP}~\cite{jinnouchiPhaseTransitionsHybrid2019, jinnouchiOntheflyMachineLearning2019}. To sample configurations, we conduct \textit{NpT} simulations at \SI{100}{\kelvin}, \SI{300}{\K}, and \SI{450}{\K} for \SI{50}{\ps} each, together with a cooling run from \SI{350}{\K} to \SI{50}{\K} over \SI{60}{\ps}. The temperature and pressure are regulated using a Parrinello-Rahman dynamics~\cite{parrinelloCrystalStructurePair1980, parrinelloPolymorphicTransitionsSingle1981}. This training procedure yields 972, 1023, and 925 structures for \smbapb{}, \smbasn{}, and \smbamixhalf{}, respectively. These training sets are combined into a unified training set of 2920 structures, enabling the MLFF to describe arbitrary metal compositions of \smbamix{} (0 $\leq$ x $\leq$ 1).

\subsection*{Molecular dynamics}

Molecular dynamics (MD) simulations are performed in the \textit{NpT} ensemble using 3×3×1 supercells. Temperature and pressure are controlled via Parrinello-Rahman dynamics, with friction coefficients of \SI{5}{\per\ps} for both atomic and lattice degrees of freedom. A time step of $\Delta t = \SI{1}{\fs}$. For robust sampling, five independent \SI{110}{\ps} runs are conducted at each temperature, with the first \SI{10}{\ps} discarded for equilibration. Trajectories are sampled every \SI{0.1}{\ps}, yielding 5000 frames per temperature for analysis.

\subsection*{Phonon calculations}

Phonons are calculated with the MLFF using the supercell method as implemented in \texttt{phonopy}~\cite{togoFirstprinciplesPhononCalculations2023, togoImplementationStrategiesPhonopy2023}. A force convergence criterion of \SI{1E-3}{\eV\per\angstrom} is used to optimize the perovskite crystal structures with the MLFF, allowing all ionic positions, cell shape, and cell volume to change. The interatomic force constants are calculated using displacements of \SI{0.01}{\angstrom} in a $3 \times 3 \times 1$ supercell. The phonon density of states (DOS) is computed on a $21 \times 21 \times 7$ $q$-mesh.

\section*{Results and discussion}

\subsection*{Chirality in static structures}

\begin{figure}[htbp]
    \includegraphics{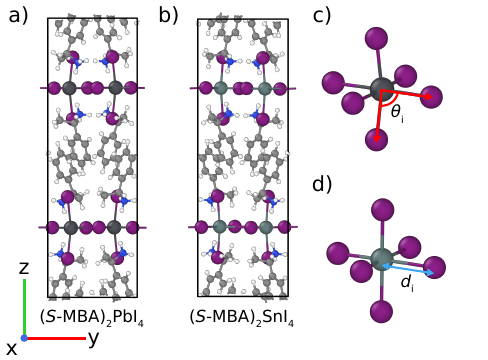}
    \caption{Structural descriptors for octahedral distortions in \mbamix{}. Unit cells of chiral (a) \smbapb{} and (b) \smbasn{}. (c) \ch{PbI6} octahedron showing a \ch{X}$-$\ch{M}$-$\ch{X} angle ($\theta_{i}$) used to compute the bond angle distortion $\sigmatwo$. (d) \ch{SnI6} octahedron showing an \ch{M}$-$\ch{X} bond length ($d_{i}$) used to compute the bond length distortion $\deltad$.} 
    \label{fig:structural_descriptors}
\end{figure}

First, we investigate the effects that different metal cations, i.e. \ch{Pb} or \ch{Sn}, have on the structural distortions and structural chirality of 2D perovskites. To do so, we analyze \mbapb{} (Figure~\ref{fig:structural_descriptors}a) and \mbasn{} (Figure~\ref{fig:structural_descriptors}b) chiral 2D perovskites, optimized using DFT calculations, the structural properties are shown in Note 1 in the Supplemental Material (SM)~\cite{supplementalMaterial} (see also references~\cite{bartokRepresentingChemicalEnvironments2013, 2dynamicsDatabase2025, liRemarkablyWeakAnisotropy2021, acharyyaIntrinsicallyUltralowThermal2020} therein). Central to this analysis are structural descriptors, which we use to capture the various types of structural distortions. All structural descriptors were computed using the \texttt{ChiraliPy} package~\cite{chiralipyCode2025}. Among these, we focus especially on structural chirality and symmetry breaking, for which we employ the descriptor set developed in our previous work on chiral 2D perovskites~\cite{polsTemperatureDependentChiralityHalide2024}.

Drawing an analogy to spin systems, We define a structural vector chirality $\bm{\epsilon}$ as
\begin{equation}
    \label{eq:vector_chirality}
    \vectorchirality = \frac{1}{N} \sum^{N}_{i = 1} \hat{\mathbf{u}}_{i} \times \hat{\mathbf{u}}_{i + 1},
\end{equation}
where $\hat{\mathbf{u}}_{i}$ are unit vectors pointing along the bonds between neighboring atoms. To measure handedness in a specific direction $\hat{\mathbf{p}}$, we project $\vectorchirality$ along that direction as $\chirality = \vectorchirality \cdot \hat{\mathbf{p}}$. A nonzero value of $\chirality$ indicates a net handedness, with the sign making a distinction between left and right. By selecting which bonds are included in Eqn.~\ref{eq:vector_chirality}, we can characterize the chirality of different components of the crystal lattice. For example, using the orientation vectors of the organic cations allows us to determine the chirality of their arrangement ($\epsA$), while including in-plane and out-of-plane metal–halide bonds yields the in-plane ($\epsMXpar$) and out-of-plane ($\epsMXperp$) chirality of the inorganic layers, respectively. Full details are given in Ref.~\cite{polsTemperatureDependentChiralityHalide2024}.

Besides these chiral descriptors, we introduce a couple of additional structural descriptors. To characterize the asymmetry in hydrogen bonding between the organic cations with the metal halide framework, we define the descriptor $\hbasym$~\cite{polsTemperatureDependentChiralityHalide2024}. For the octahedral distortions, we use the descriptors introduced by \citet{robinsonQuadraticElongationQuantitative1971}. The relative spread in metal halide bond lengths of a \ch{MX6} octahedron is defined by
\begin{equation}
    \label{eq:bond_length_variance}
    \deltad = \frac{1}{6} \sum_{i = 1}^{6} \left( \frac{d_{i} - d_{0}}{d_{0}} \right)^{2},
\end{equation}
where $d_{i}$ is the bond length of one of the \ch{M}$-$\ch{X} bonds (Figure~\ref{fig:structural_descriptors}c) and $d_{0}$ the average length of all \ch{M}$-$\ch{X} bonds in an octahedron. The bond angle variance $\sigmatwo$ is computed as
\begin{equation}
    \label{eq:bond_angle_variance}
    \sigmatwo = \frac{1}{11} \sum_{i = i}^{12} \left( \theta_{i} - \theta_{0} \right)^{2},
\end{equation}
with $\theta_{i}$ one of the twelve \textit{cis} \ch{X}$-$\ch{M}$-$\ch{X} angles (Figure~\ref{fig:structural_descriptors}d) and $\theta_{0} =$ \SI{90}{\degree} the value for such bond angles in an ideal octahedron. 

\begin{table*}[htbp]
    \caption{Structural descriptors probing structural chirality and bond asymmetry in chiral 2D perovskite structures. \ch{Pb}-based enantiomers were mirrored, yielding chiral descriptor values of equal magnitude, while \ch{Sn}-based enantiomers used two independent experimental structures for \textit{S}- and \textit{R}-enantiomers, causing slight variations in the descriptor magnitudes.}
    \label{tab:structural_descriptors} 
    \begin{ruledtabular}
    \begin{tabular}{cccccccc}
        Perovskite                      & Ref.                                                  & $\epsA$  ($\times$10\textsuperscript{-3})     & $\epsMXpar$ ($\times$10\textsuperscript{-3})      & $\epsMXperp$ ($\times$10\textsuperscript{-3})     & $\hbasym$  (\SI{}{\angstrom})     & $\deltad$ ($\times$10\textsuperscript{-4})    & $\sigmatwo$ (\SI{}{\degree\squared})      \\ \hline
        \smbapb{}                       & \cite{janaOrganictoinorganicStructuralChirality2020}  & +46.490                                       & +4.534                                            & +1.560                                            & 0.036                             & 2.938                                         & 20.17                                     \\
        \rmbapb{}                       & \cite{janaOrganictoinorganicStructuralChirality2020}  & -46.490                                       & -4.534                                            & -1.560                                            & 0.036                             & 2.938                                         & 20.17                                     \\
        \racmbapb{}                     & \cite{dangBulkChiralHalide2020}                       & 0.000                                         & 0.000                                             & 0.000                                             & 0.000                             & 1.819                                         & 14.04                                     \\ \hline
        \smbasn{}                       & \cite{luHighlyDistortedChiral2020}                    & +44.061                                       & +5.816                                            & +0.792                                            & 0.039                             & 75.055                                        & 7.98                                      \\
        \rmbasn{}                       & \cite{luHighlyDistortedChiral2020}                    & -44.336                                       & -5.543                                            & -0.570                                            & 0.040                             & 75.742                                        & 7.76                                      \\
        \racmbasn{}                     & \cite{luHighlyDistortedChiral2020}                    & 0.000                                         & 0.000                                             & 0.000                                             & 0.000                             & 65.464                                        & 13.66                                     \\
    \end{tabular}
    \end{ruledtabular}
\end{table*}

The values of all descriptors as computed for the studied \ch{Pb}- and \ch{Sn}-based perovskites in their optimized equilibrium structures can be found in Table~\ref{tab:structural_descriptors}. The similarity of the crystal structures of \smbapb{} and \smbasn{} enables a direct comparison their structural descriptors. Notably, the chiral descriptors of both compounds are of similar magnitude. This holds for the arrangement of the organic cations ($\epsA$) and the in-plane chirality of the metal halide layer ($\epsMXpar$). The only marked difference is observed in the out-of-plane chirality ($\epsMXperp$), which is approximately twice as large in the \ch{Pb} compound compared to the \ch{Sn} analogue. However, the overall magnitude of $\epsMXperp$ remains relatively small. In addition, the asymmetry in hydrogen bonding between the cations and the framework ($\hbasym$) is also similar in both materials.

The similarity of the values for the chiral descriptors is in stark contrast with the pattern observed for the descriptors for octahedral distortions. In \ch{Sn}-based perovskites, the in-plane \ch{M}$-$\ch{X} bonds show significantly larger variations in the bond lengths compared to those in \ch{Pb}-based compounds (SM Note 2~\cite{supplementalMaterial}), as reflected by the considerably higher value of $\deltad$ for \mbasn{}~\cite{luHighlyDistortedChiral2020}. By comparison, we find that $\sigmatwo$ is larger in \ch{Pb}-based perovskites. This increase in $\sigmatwo$ is foremost due to larger deviations of the axial \ch{X}$-$\ch{M}$-$\ch{X} angles away from \SI{90}{\degree}, whereas the equatorial angles display a narrower distribution around \SI{90}{\degree}. These observations align with the nearly identical values of $\epsMXpar$ in both \ch{Pb}- and \ch{Sn}-based compounds, while $\epsMXperp$ shows a large difference.

The noticeably larger bond length distortions in the \ch{Sn} halide octahedra are due to an off-centering of the \ch{Sn} atom, driven by the stereochemically active 5s$^{2}$ lone pair of the \ch{Sn^{2+}} ion. The 5s-5p splitting on \ch{Sn^{2+}} is relatively small~\cite{taoAbsoluteEnergyLevel2019}, which allows for an sp$^{n}$ hybridization that enables an asymmetric (non-spherical) lone pair electron density and concomitant structural distortion of the \ch{SnI6} octahedra~\cite{fabiniDynamicStereochemicalActivity2016, remsingNewPerspectiveLone2020, hylton-farringtonOctahedralTiltingBsite2025}. In contrast, the 6s$^{2}$ lone pair on \ch{Pb^{2+}} is stereochemically significantly less active because of its deeper energy level~\cite{taoAbsoluteEnergyLevel2019}. Consequently, there is little to no hybridization between the \ch{Pb} 6s and 6p orbitals, preserving the centrosymmetry of the \ch{Pb}-based octahedra. Alternatively, we note that recently hard and soft acid-base (HSAB) theory has been applied to argue that \ch{SnI6} octahedra are more easily deformed than \ch{PbI6} octahedra~\cite{fortinoRoleMetalhalideBond2024}. Ultimately, it is somewhat remarkable that, despite the pronounced octahedral distortions in the \ch{Sn}-based compounds, their structural chirality remains largely unaffected.

\subsection*{Finite-temperature chirality}

Next, we assess the finite temperature dynamics of 2D halide perovskites with mixed metal compositions, such as \mbamix{}. We use the MLFF trained against the unified training set with various metal compositions, the full details of the training procedure can be found in SM Note 3~\cite{supplementalMaterial}. Validations show that our model has high accuracy and transferability across various metal compositions and crystal structures (e.g. \textit{S}-, \textit{R}-, and \textit{rac}-structures), predicting energies, forces, and stresses with a low error compared to DFT calculations.

\begin{figure*}[htbp]
    \includegraphics{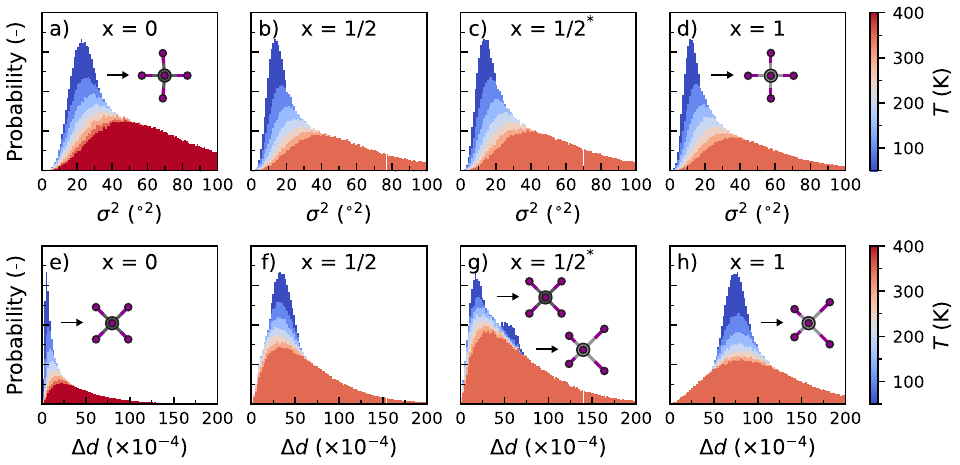}
    \caption{Finite temperature octahedral distortions in \smbamix{}. Bond angle variance $\sigmatwo$ of (a) \smbapb{} (x $=$ 0), (b) ordered mixed and (c) randomly mixed \smbamixhalf{} (x $=$ 1/2), and (d) \smbasn{} (x $=$ 1). Bond length variance $\deltad$ of (e) \smbapb{} (x $=$ 0), (f) ordered mixed and (g) randomly mixed \smbamixhalf{} (x $=$ 1/2), and (h) \smbasn{} (x $=$ 1). The insets show (a-d) side view and (e-h) top view of the metal halide octahedra.}
    \label{fig:octahedral_distortions}
\end{figure*}

\begin{figure}[htbp]
    \includegraphics{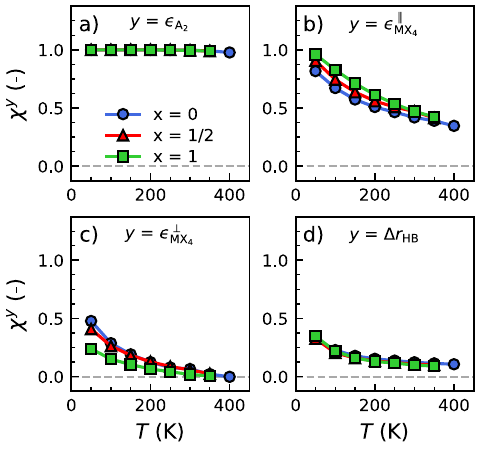}
    \caption{Temperature dependence of degree of chirality in \smbamix{}. The values of $\degreechi$ are shown for (a) $\epsA$, (b) $\epsMXpar$, (c) $\epsMXperp$, and (d) $\hbasym$. The investigated perovskites have different metal cation compositions including \smbapb{} (x $=$ 0), ordered mixed \smbamix{} (x $=$ 1/2), and \smbasn{} (x $=$ 1).}
    \label{fig:temperature_dependence}
\end{figure}

We use this MLFF in molecular dynamics (MD) simulations of \smbamix{} with different compositions (x $=$ 0, 1/2, and 1). Additional details on how the systems with a mixed composition (x = 1/2) are generated can be found in SM Note 4~\cite{supplementalMaterial}. The simulations are carried out at various temperatures (50 - \SI{400}{\K}) in the \textit{NpT} ensemble at atmospheric pressure. During these simulations we monitor the atomic structure and the finite temperature dynamics of the octahedral distortions as captured by $\sigmatwo$ and $\deltad$ (Figure~\ref{fig:octahedral_distortions}) and the structural chirality descriptors (Figure~\ref{fig:temperature_dependence}).

Focusing on the octahedral distortions and their dynamics, shown in Figure~\ref{fig:octahedral_distortions}, we find that the same trends observed in static structures are maintained at finite temperatures. The bond angle variance $\sigmatwo$ is greater in \smbapb{} (Figure~\ref{fig:octahedral_distortions}a-d), and the bond length variance $\deltad$ is significantly larger in \smbasn{} (Figure~\ref{fig:octahedral_distortions}e-h). In all perovskites, the $\sigmatwo$ and $\deltad$ distributions are asymmetric, with the distributions shifting to higher values for increasing temperatures.

An ordered checkerboard mixing of the metal cations modifies the octahedral distortions. While the $\sigmatwo$ remains similar to that of the \ch{Sn}-based perovskites (Figure~\ref{fig:octahedral_distortions}b), the bond length variance $\deltad$ becomes an average of the two pure systems (Figure~\ref{fig:octahedral_distortions}f). Random mixing of the metal cations results in the formation of \ch{Pb}- and \ch{Sn}-rich octahedra. This does not seem to affect $\sigmatwo$ much (Figure~\ref{fig:octahedral_distortions}c), but $\deltad$ now adopts a bimodal distribution (Figure~\ref{fig:octahedral_distortions}g). The low $\deltad$ peak, which closely resembles \smbapb{}, corresponds to \ch{Pb}-rich domains, while the high $\deltad$ peak is associated with \ch{Sn}-rich domains. The bimodal distribution of the bond length variance is most pronounced at low temperatures ($<$ \SI{150}{\K}), and is washed out at higher temperatures.

Next, we investigate the chirality of the halide perovskites. To do so, we use the degree of chirality $\degreechi$, introduced in our earlier work~\cite{polsTemperatureDependentChiralityHalide2024}, which is defined as
\begin{equation}
    \label{eq:degree_of_chirality}
    \degreechi = \frac{\phiS - \phiR}{\phiS + \phiR}
\end{equation}
where $\phiS$ and $\phiR$ are the fractions of the distribution of a particular chiral descriptor, i.e. $y = \epsA$, $\epsMXpar$, $\epsMXperp$, or $\hbasym$, that we associate with the \textit{S}- and \textit{R}-enantiomer, respectively. The degree of chirality can attain values $-1$ $\leq$ $\degreechi$ $\leq$ $+1$, with $+1$ ($-1$) representing the perfect \textit{S}-enantiomer (\textit{R}-enantiomer) and zero an achiral structure. Structures with reduced chirality have values $0 < |\degreechi| < 1$.

In Figure~\ref{fig:temperature_dependence}, we show the temperature dependence of $\degreechi$ for the various components of the \smbamix{} perovskite. At low temperature (\SI{50}{\K}), the cation arrangement ($\epsA$) is the most chiral across all compositions, with $\degreechi$ $=$ +1.00. The in-plane inorganic distortions ($\epsMXpar$) are also highly chiral, with $\degreechi$ between +0.82 and +0.96. The other components have a substantially lower degree of chirality between +0.24 and +0.48 for the out-of-plane inorganic framework distortions ($\epsMXperp$) and $\degreechi$ between +0.33 and +0.35 for the hydrogen bond asymmetry ($\hbasym$). Notably, we find that the trends in the magnitude of the structural descriptors (Table~\ref{tab:structural_descriptors}) are preserved in the degree of chirality. To demonstrate, \smbasn{} (x $=$ 1) exhibited the largest in-plane chirality in the inorganic layers, and thus it exhibits the largest degree of chirality for $\epsMXpar$. Similarly, \smbapb{} (x $=$ 0), with its large out-of-plane chirality, has the largest degree of chirality for $\epsMXperp$.

With increasing temperature, $\degreechi$ decreases for all descriptors except for the organic cation arrangement ($\epsA$), whose chirality is preserved due to the locked orientation of the large organic cations in their packing (Figure~\ref{fig:temperature_dependence}a). In contrast, the inorganic framework loses its chirality, $\epsMXpar$, $\epsMXperp$, with increasing temperature (Figure~\ref{fig:temperature_dependence}b-c). Increased thermal motion leads to a less persistent hydrogen bonding to the organic cations (Figure~\ref{fig:temperature_dependence}d), and as this is the mechanism by which the inorganic framework acquires chirality~\cite{sonUnravelingChiralityTransfer2023, polsTemperatureDependentChiralityHalide2024}, the latter decreases with increasing temperature. 

When different compositions are compared, we find that at \SI{300}{\K} and below, the different perovskite compositions result in a varying degree of chirality. These difference may affect the lattice vibrations of these materials. Interestingly, at higher temperatures ($\geq$ \SI{350}{\K}), the chirality of the structures converges to similar values, regardless of the metal composition, indicating that temperature effects override structural differences. Furthermore, the mixing of the metal cations, ordered or random, does not impact the temperature dependence of the chirality (SM Note 5~\cite{supplementalMaterial}). The similarity in this decay can be related to the observed similarities in the hydrogen bonds across various compositions and atomic details of the mixing. This is evidenced by similar orientational lifetimes for the \ch{NH3^{+}} head groups, connecting the organic and inorganic components of the perovskites  (SM Note 6~\cite{supplementalMaterial}).

\subsection*{Chiral lattice vibrations}

We shift our focus to the phonon spectra of chiral perovskites. Using the MLFF trained for different metal compositions, we calculate the vibrational properties of \smbapb{} and \smbasn{}. More details of these phonon calculations can be found in SM Note 7~\cite{supplementalMaterial}.

\begin{figure*}[htbp!]
    \includegraphics{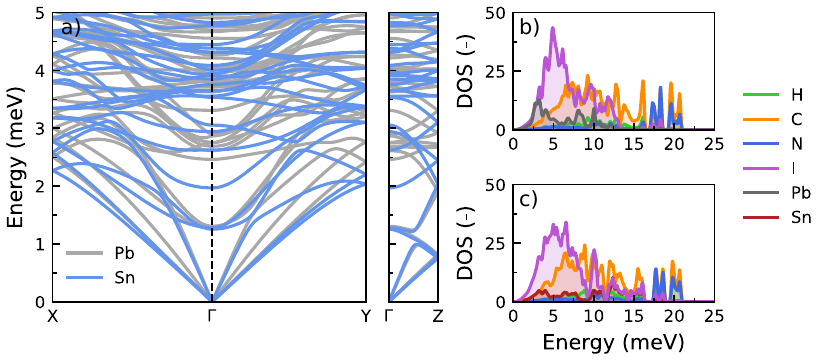}
    \caption{Phonons in chiral 2D perovskites. (a) Phonon dispersions with Brillouin zone is sampled along high-symmetry paths Γ$-$X, Γ$-$Y, and Γ$-$Z. The dispersion of \smbapb{} is shown in gray and that of \smbasn{} is shown in blue. Phonon density of states (DOS) of the low energy region (0 - \SI{25}{\meV}) of (b) \smbapb{} and (c) \smbasn{}. A Gaussian broadening of \SI{0.1}{\meV} is used for the DOS.}
    \label{fig:phonon_dispersion_dos}
\end{figure*}

In Figure~\ref{fig:phonon_dispersion_dos}, we show the phonon spectra of chiral \smbapb{} and \smbasn{}. The dispersion of the phonons is shown for the in-plane (Γ$-$X and Γ$-$Y) and out-of-plane (Γ$-$Z) directions (Figure~\ref{fig:phonon_dispersion_dos}a). Both perovskites show similar phonon dispersion features, but the \ch{Sn}-based perovskite has lower energies for the phonon branches below \SI{3.5}{\meV}. These energy shifts are very apparent for the acoustic phonons in the in-plane direction, as reflected by differences in the group velocities of phonons propagating along the in-plane directions, the details of which are found in SM Note 7~\cite{supplementalMaterial}. In particular, the in-plane acoustic phonons in \smbasn{} propagate with lower group velocities (\SI{1587}{\m\per\s} and \SI{1637}{\m\per\s}) than those in \smbapb{} (\SI{1831}{\m\per\s} and \SI{1866}{\m\per\s}). This may seem unexpected, given that \ch{Pb} is 75\% heavier than \ch{Sn}. Nevertheless, it suggests that the ions in the \ch{Sn}$-$\ch{I} network is somewhat less strongly bonded than the \ch{Pb}$-$\ch{I} network (see SM Note 7~\cite{supplementalMaterial}).

The largest differences between the Sn- and Pb-based compounds appear in the low-energy region of the phonon spectrum. Overall, however, their phonon densities of states (DOS) are very similar (Figure~\ref{fig:phonon_dispersion_dos}b-c). Across the full range of vibrational modes in the chiral perovskites (Figure S8), both materials exhibit comparable behavior; the low-energy modes (0$-$\SI{25}{\meV}) originate from the inorganic framework, while the high-energy modes ($>$ \SI{25}{\meV}) stem from vibrations of the organic cations.

Next, we characterize the nature of the phonons in chiral 2D perovskites. Specifically, given the presence of chiral phonons in chiral 2D perovskites~\cite{polsChiralPhonons2D2025}, we probe the circular polarization or chirality of the phonons ($\phononcircpolgeneral$). To do so, we calculate the phonon angular momentum of an eigenmode at wave vector $\mathbf{q}$ and mode index $\sigma$ as
\begin{equation}
    \phononcircpolgeneral = \sum^{N}_{i = 1} \mathbf{e}^{\dagger}_{i, \mathbf{q}, \sigma} \mathbf{S}^{\alpha} \mathbf{e}_{i, \mathbf{q}, \sigma},
\end{equation}
where $\mathbf{e}_{i, \mathbf{q}, \sigma}$ is the polarization vector of the $i$\textsuperscript{th} atom in the unit cell, $\mathbf{S}^{\alpha}$ $(\alpha = x, y, z$) are the spin-1 matrices on a Cartesian basis, and $N$ is the total number of atoms in the unit cell. The magnitude and sign of this phonon angular momentum determine the chirality or handedness of the phonon, with 1 $\geq$ $\phononcircpolgeneral$ $>$ $0$, $-1$ $\leq$ $\phononcircpolgeneral$ $<$ $0$, and $\phononcircpolgeneral$ $=$ 0 indicating a right-handed, left-handed, and achiral phonon mode, respectively. More details of this procedure can be found in Refs.~\cite{zhangChiralPhononsHighSymmetry2015, chenChiralPhononDiode2022, polsChiralPhonons2D2025}.

\begin{figure*}[htbp]
    \includegraphics{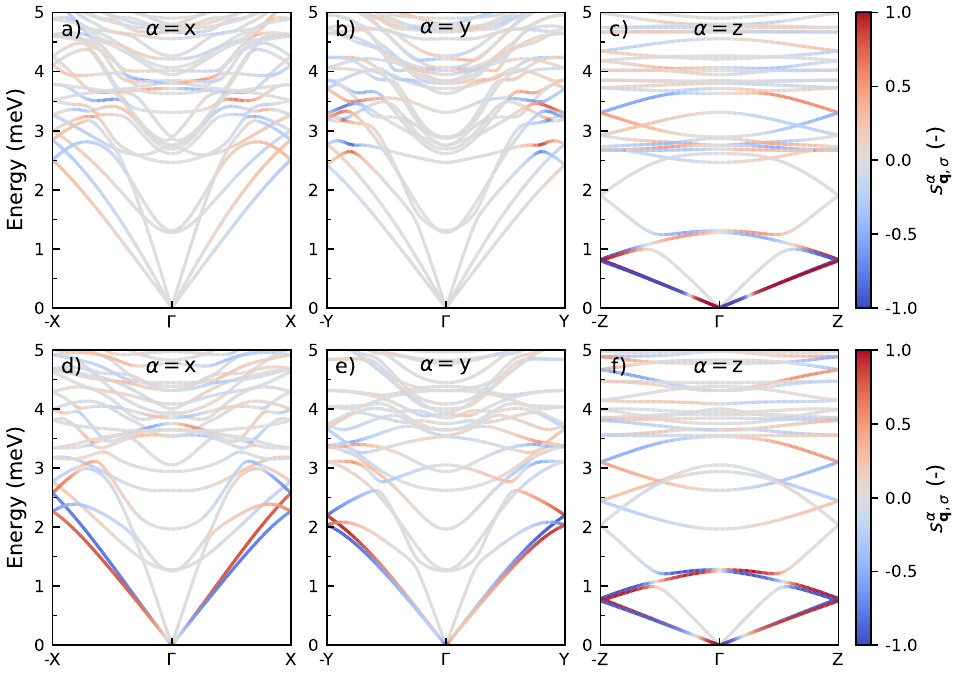}
    \caption{Circularly polarized phonon dispersions of (a-c) \smbapb{} and (d-f) \smbasn{}. Phonon branches are colored according to their circular polarization. Red, blue and gray represent right-handed ($\phononcircpolgeneral$ $>$ 0), left-handed ($\phononcircpolgeneral$ $<$ 0), and non-polarized ($\phononcircpolgeneral$ $=$ 0) phonon modes.}
    \label{fig:chiral_phonon_spectrum}
\end{figure*}

Figure~\ref{fig:chiral_phonon_spectrum} presents the circularly polarized phonon dispersion of \smbapb{} and \smbasn{} for phonons propagating along the $x$- (Γ$-$X), $y$- (Γ$-$Y) and $z$-directions (Γ$-$Z). Consistent with a previous study on phonons in chiral perovskites, the phonons are polarized around their propagation axis along high-symmetry paths, a consequence of the \triplescrew{} space group. For instance, the phonons propagating along the $z$-axis (Γ$-$Z) exhibit circular motion in the $xy$-plane. Comparing the chirality of phonons in \ch{Pb}- and \ch{Sn}-based perovskites reveals a similar circular polarization. Specifically, the lowest two acoustic phonon branches in the in-plane direction share a similar chirality in both compositions; left- and right-handed along the Γ$-$X path, and right- and left-handed along the Γ$-$Y path. This similarity arises from their nearly identical crystal structures.

The in-plane acoustic modes show a marked increase in chirality in \smbasn{} compared to \smbapb{}. This correlates with the structural chirality along the in-plane direction $\epsMXpar$ in Table~\ref{tab:structural_descriptors}, which is also larger in \smbasn{} compared to \smbapb{}. A structure-property relationship might exist between structural chirality and phonon chirality, which would suggest a pathway for tuning phonon chirality through structural modifications. In contrast, the perovskite composition appears to have only a minor effect on the chirality of acoustic phonons propagating along the out-of-plane direction (Γ$-$Z). Additionally, the near-degeneracy of the low-energy acoustic branches allows for phonon pairs to have arbitrary polarizations, making it difficult to draw conclusions about their chirality. When the metal cations are mixed, \smbamixhalf{}), the resulting phonon dispersions reflect averages of those observed in the pure compounds, with the associated phonon chirality having intermediate values, as shown in SM Note 7~\cite{supplementalMaterial}.

Chiral phonons can be observed in experiments through heat transport experiments. As predicted by \citet{hamadaPhononAngularMomentum2018}, a flux of chiral phonons, created by driving a phonon distribution out of equilibrium through a temperature gradient, can generate a finite amount of angular momentum. The exact proportionality between the components of the phonon angular momentum per unit cell volume $J^{\mathrm{ph},\alpha}$ and the temperature gradient $\partial T / \partial x^{\beta}$ is given by
\begin{equation}
    \label{eq:volumetric_angular_momentum_generation}
    \begin{split}
    J^{\mathrm{ph},\alpha} &= - \frac{\hbar\tau}{V} \sum_{\mathbf{q}, \sigma; \beta=x,y,z} s^{\alpha}_{\mathbf{q}, \sigma} v^\beta_{\mathbf{q}, \sigma} \frac{\partial f_{0} \left( \omega_{\mathbf{q}, \sigma} \right)}{\partial T} \frac{\partial T}{\partial x^{\beta}} \\
    &\equiv \sum_{\beta} \alpha^{\alpha \beta} \frac{\partial T}{\partial x^{\beta}},
    \end{split}
\end{equation}
where $V$ is the unit cell volume, $s^{\alpha}_{\mathbf{q}, \sigma}$ the phonon angular momentum, $v^\beta_{\mathbf{q}, \sigma}$ the phonon group velocity, and $f_{0}$ the Bose-Einstein distribution. Eqn.~\ref{eq:volumetric_angular_momentum_generation} holds in the uniform relaxation time approximation, with $\tau$ the phonon relaxation time. The response tensor is $\alpha^{\alpha \beta}$, the components of which are given in superscripts ($\alpha, \beta = x, y, z$). For the \triplescrew{} space group this tensor is purely diagonal, i.e. $\alpha^{\alpha \beta} = 0$ ($\alpha \neq \beta$), with unequal diagonal elements $\alpha^{xx} \neq \alpha^{yy} \neq \alpha^{zz}$.

\begin{figure*}[htbp]
    \includegraphics{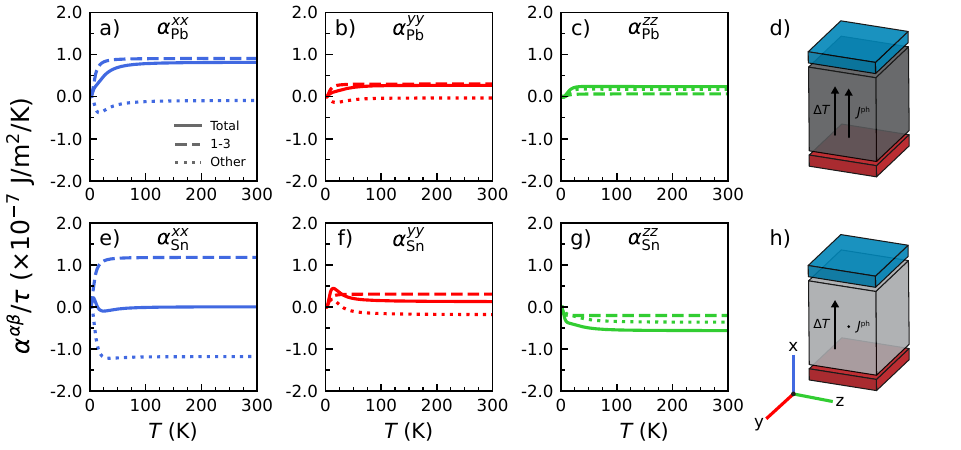}
    \caption{Generated angular momentum due to chiral phonons. The temperature dependence of the (a) $\alpha^{xx}$, (b) $\alpha^{yy}$, and (c) $\alpha^{zz}$ response tensors for \smbapb{}, with (d) a schematic overview showing its response in the $x$-direction. The temperature dependence of the (e) $\alpha^{xx}$, (f) $\alpha^{yy}$, and (g) $\alpha^{zz}$ response tensors for \smbasn{}, with (h) a schematic overview showing its response in the $x$-direction.}
    \label{fig:angular_momentum_generation}
\end{figure*}

In Figure~\ref{fig:angular_momentum_generation} we show the temperature dependence of the response tensor for \smbapb{} and \smbasn{}. Since all curves level off at around \SI{150}{\K}, this confirms that it is primarily the low energy phonons that contribute to the effect. Examining the response tensors at \SI{300}{\K}, we find that the \ch{Pb}-based compound (Figure~\ref{fig:angular_momentum_generation}a-c) appears to only generate an appreciable amount of angular momentum in the $x$-direction ($\alpha^{xx}_{\ch{Pb}} \gg \alpha^{yy}_{\ch{Pb}} \approx \alpha^{zz}_{\ch{Pb}}$). In contrast, the \ch{Sn}-based compound (Figure~\ref{fig:angular_momentum_generation}e-g) generates substantially smaller amounts of angular momentum in the $x$-direction ($\alpha^{xx}_{\ch{Pb}} \gg \alpha^{xx}_{\ch{Sn}}$), with, like the Pb-based compound, the $y$- and $z$-directions also not generating large amounts of angular momentum ($\alpha^{xx}_{\ch{Sn}} \approx \alpha^{yy}_{\ch{Sn}} \approx \alpha^{zz}_{\ch{Sn}}$). We highlight that the amount of angular momentum generated in \smbapb{} is large enough to be experimentally observable through macroscopic rigid-body rotation of small crystals, whereas it is not in \smbasn{} as detailed in SM Note 8~\cite{supplementalMaterial}.

The difference in the $\alpha^{xx}$ response tensor of \smbapb{} and \smbasn{} arises from different contributions of the various parts of the phonon spectrum. For \smbapb{}, the main part of the response $\alpha^{xx}$ originates from the three lowest frequency branches of phonons, the other branches contribute very little (Figure~\ref{fig:angular_momentum_generation}a). As a result, the response of the \ch{Pb}-based perovskite is dominated by the low-energy phonon branches. For \smbasn{} the group of lowest three frequency branches and the group of slightly higher energy phonon branches each provide a substantial contribution to $\alpha^{xx}$. However, the contributions of the two groups have opposite sign (Figure~\ref{fig:angular_momentum_generation}e), and largely compensate one another. The net result is that the \ch{Sn}-based perovskite generates a negligible amount of angular momentum under a temperature gradient.

\section*{Conclusion}

In conclusion, we study the effects that the perovskite composition has on the chirality of 2D metal halide perovskites. To do so, we investigate \mbamix{} perovskites with varying metal composition (x = 0, 1/2, and 1) and in the case of mixed metal systems the type of mixing, i.e. ordered and random. Both \ch{Sn^{2+}} and \ch{Pb^{2+}} cations have a lone pair (5s$^2$ for \ch{Sn^{2+}} and 6s$^2$ for \ch{Pb^{2+}}), but the one on tin is significantly more stereoactive. This results in a structural distortion of tin halide octahedra, which manifests in \ch{Sn^{2+}} being off-centered. The octahedral asymmetry has a moderate effect on the structural chirality, which for \smbasn{} and \smbapb{} is quite similar. Moreover, at high temperatures ($\geq$ \SI{350}{\K}), the increase in ionic motion makes all structural differences negligible.

Focusing on the lattice vibrations, both \smbapb{} and \smbasn{} exhibit similar phonon spectra, which reflects their similar structures and bonding patterns. The distorted nature of the octahedra in \ch{Sn}-based perovskites weakens the coupling between the octahedra, decreasing the phonon group velocity for the in-plane acoustic phonons compared to \ch{Pb}-based perovskites. The somewhat larger in-plane chirality in \smbasn{} coincides with a much larger chirality of the in-plane acoustic phonons when compared to \smbapb{}. Despite the larger chirality of the in-plane phonons for \ch{Sn}-based perovskites, the angular momentum generated due to a temperature gradient is smaller than in \ch{Pb}-based perovskites, as a result of a compensating effect of the higher energy phonon branches in \smbasn{}. Altogether, this highlights the interplay between the structural distortions, structural chirality and phonon chirality. 

\section*{Data availability}

The training sets for the machine-learning force fields (MLFFs) are available at Ref.~\cite{2dynamicsDatabase2025}. The analysis script for the structural descriptors and the resulting processed data are available at Ref.~\cite{chiralipyCode2025}. The data and scripts used to create the figures in the manuscript are available at Ref.~\cite{zenodoData2025}. Additional data are available from the corresponding authors upon reasonable request.

\begin{acknowledgments}

S.T. acknowledges funding from Vidi (Project VI.Vidi.213.091) from the Dutch Research Council (NWO).

\end{acknowledgments}

\bibliography{ms}

\end{document}


\title{Supplemental Material: \\ Impact of Metal Cation on Chiral Properties of 2D Halide Perovskites}

\author{Mike Pols}
\email{m.c.w.m.pols@tue.nl}
\affiliation{%
    Materials Simulation \& Modelling, Department of Applied Physics and Science Education, Eindhoven University of Technology, 5600 MB, Eindhoven, The Netherlands
}%

\author{Helena Boom}
\affiliation{%
    Materials Simulation \& Modelling, Department of Applied Physics and Science Education, Eindhoven University of Technology, 5600 MB, Eindhoven, The Netherlands
}%

\author{Geert Brocks}
\affiliation{%
    Materials Simulation \& Modelling, Department of Applied Physics  and Science Education, Eindhoven University of Technology, 5600 MB, Eindhoven, The Netherlands
}%
\affiliation{%
    Computational Chemical Physics, Faculty of Science and Technology and MESA+ Institute for Nanotechnology, University of Twente, 7500 AE, Enschede, The Netherlands
}%

\author{Sof\'{i}a Calero}
\affiliation{%
    Materials Simulation \& Modelling, Department of Applied Physics and Science Education, Eindhoven University of Technology, 5600 MB, Eindhoven, The Netherlands
}%

\author{Shuxia Tao}
\email{s.x.tao@tue.nl}
\affiliation{%
    Materials Simulation \& Modelling, Department of Applied Physics and Science Education, Eindhoven University of Technology, 5600 MB, Eindhoven, The Netherlands
}%

\keywords{Metal halide perovskites, mixing, metal cations, chirality, structural chirality, phonon chirality, angular momentum}

\maketitle

\clearpage

\tableofcontents

\clearpage

\section{Density functional theory (DFT) calculations}

An overview of the experimental crystal structures that are optimized is provided in Table~\ref{tab:perovskite_structures}. The resulting optimized geometries and the $k$-meshes used are shown in Table~\ref{tab:k-meshes}.

\begin{table}[htbp]
    \caption{Experimental crystal structures of investigated 2D halide perovskites.}
    \label{tab:perovskite_structures}
    \begin{ruledtabular}
    \begin{tabular}{cccccc}
        Perovskite                  & Space group                       & $N_{\mathrm{group}}$  & $T_{\mathrm{exp.}}$ (\SI{}{\K})   & Database ID                          & References                                                                    \\ \hline
        \smbapb{}                   & \triplescrew                      & 19                    & 298                               & \ccdc{2015617}                       & Ref.\citep{janaOrganictoinorganicStructuralChirality2020}                     \\
        \rmbapb{}                   & \triplescrew                      & 19                    & 298                               & \ccdc{2015617} (mirror)              & Ref.\citep{janaOrganictoinorganicStructuralChirality2020}                     \\
        \racmbapb{}                 & \centroscrew                      & 14                    & 293                               & \ccdc{1877052}                       & Ref.\citep{dangBulkChiralHalide2020}                                          \\
        \smbasn{}                   & \triplescrew                      & 19                    & 250                               & \ccdc{1994338}                       & Ref.\citep{luHighlyDistortedChiral2020}                                       \\
        \rmbasn{}                   & \triplescrew                      & 19                    & 250                               & \ccdc{1994337}                       & Ref.\citep{luHighlyDistortedChiral2020}                                       \\
        \racmbasn{}                 & \pnma                             & 19                    & 250                               & \ccdc{1994336}                       & Ref.\citep{luHighlyDistortedChiral2020}                                       \\
    \end{tabular}
    \end{ruledtabular}
\end{table}

\begin{table}[htbp]
    \caption{Lattice vector lengths ($a$, $b$ and $c$) and $k$-meshes used to calculate the reference data for the machine-learning force fields (MLFFs).}
    \label{tab:k-meshes}
    \begin{ruledtabular}
    \begin{tabular}{cccccc}
        Training structure                      & XC functional         & $a$ (\SI{}{\angstrom})        & $b$ (\SI{}{\angstrom})        & $c$ (\SI{}{\angstrom})        & $k$-meshes                    \\ \hline
        \smbapb{}                               & SCAN                  &  8.87                         &  9.19                         & 28.76                         & \kpoints{2}{2}{1}             \\
        \smbamixhalf{}                          & SCAN                  &  8.90                         &  9.24                         & 28.73                         & \kpoints{2}{2}{1}             \\
        \smbasn{}                               & SCAN                  &  8.93                         &  9.28                         & 28.56                         & \kpoints{2}{2}{1}             \\ 
    \end{tabular}
    \end{ruledtabular}
\end{table}

\clearpage

\section{Bond lengths and bond angles}

To compute the octahedral distortions of the perovskites we take into account the entire geometry of a metal halide octahedron, i.e. all bond lengths $d_{i}$ and bond angles $\theta_{i}$. In Table~\ref{tab:bond_angles} and Table~\ref{tab:bond_lengths} we show the individual bond angles and bond lengths of the octahedra in these perovskites, respectively.

\begin{table}[htbp]
    \caption{Octahedral angles in 2D perovskites.}
    \label{tab:bond_angles}
    \begin{ruledtabular}
    \begin{tabular}{cccccccc}
        \multirow{3}{*}{Perovskite}     & \multicolumn{4}{c}{Axial} & \multicolumn{2}{c}{Equatorial} \\
                                        & $\theta_{1}$ (\SI{}{\degree}) & $\theta_{2}$ (\SI{}{\degree}) & $\theta_{3}$ (\SI{}{\degree}) & $\theta_{4}$ (\SI{}{\degree}) & $\theta_{9}$ (\SI{}{\degree}) & $\theta_{10}$ (\SI{}{\degree}) \\ 
                                        & $\theta_{5}$ (\SI{}{\degree}) & $\theta_{6}$ (\SI{}{\degree}) & $\theta_{7}$ (\SI{}{\degree}) & $\theta_{8}$ (\SI{}{\degree}) & $\theta_{11}$ (\SI{}{\degree}) & $\theta_{12}$ (\SI{}{\degree}) \\ \hline
        \multirow{2}{*}{\smbapb{}}      & 91.02 & 84.22 & 89.09 & 97.17 & 88.05 & 90.36 \\
                                        & 89.90 & 82.83 & 88.68 & 95.82 & 86.15 & 95.42 \\ \hline
        \multirow{2}{*}{\rmbapb{}}      & 91.02 & 84.22 & 89.09 & 97.17 & 88.05 & 90.36 \\
                                        & 89.90 & 82.83 & 88.68 & 95.82 & 86.15 & 95.42 \\ \hline
        \multirow{2}{*}{\racmbapb{}}    & 89.90 & 85.33 & 90.10 & 94.67 & 85.91 & 94.09 \\
                                        & 90.10 & 94.67 & 89.90 & 85.33 & 94.09 & 85.91 \\ \hline
        \multirow{2}{*}{\smbasn{}}      & 92.10 & 87.76 & 85.64 & 91.76 & 91.23 & 89.76 \\
                                        & 94.99 & 88.56 & 87.15 & 91.80 & 86.69 & 92.31 \\ \hline
        \multirow{2}{*}{\rmbasn{}}      & 91.93 & 87.75 & 85.86 & 91.85 & 91.03 & 89.90 \\
                                        & 94.99 & 88.53 & 87.08 & 91.76 & 86.72 & 92.35 \\ \hline
        \multirow{2}{*}{\racmbasn{}}    & 91.98 & 86.48 & 87.77 & 93.49 & 92.24 & 88.42 \\
                                        & 91.98 & 86.48 & 87.77 & 93.49 & 83.52 & 95.82 \\
    \end{tabular}
    \end{ruledtabular}
\end{table}

\begin{table}[htbp]
    \caption{Octahedral bond lengths and average bond length in 2D perovskites.}
    \label{tab:bond_lengths}
    \begin{ruledtabular}
    \begin{tabular}{cccccccc}
        \multirow{2}{*}{Perovskite} & \multicolumn{2}{c}{Axial} & \multicolumn{4}{c}{Equatorial} & Average \\ 
                                    & $d_{1}$ (\SI{}{\angstrom}) & $d_{2}$ (\SI{}{\angstrom}) & $d_{3}$ (\SI{}{\angstrom}) & $d_{4}$ (\SI{}{\angstrom}) & $d_{5}$ (\SI{}{\angstrom}) & $d_{6}$ (\SI{}{\angstrom}) & $d_{0}$ (\SI{}{\angstrom}) \\ \hline
        \smbapb{}                   & 3.190 & 3.211 & 3.213 & 3.324 & 3.315 & 3.305 & 3.260 \\
        \rmbapb{}                   & 3.190 & 3.211 & 3.213 & 3.324 & 3.315 & 3.305 & 3.260 \\
        \racmbapb{}                 & 3.200 & 3.200 & 3.238 & 3.306 & 3.238 & 3.306 & 3.248 \\ \hline
        \smbasn{}                   & 3.190 & 3.148 & 3.738 & 3.559 & 3.008 & 2.973 & 3.269 \\
        \rmbasn{}                   & 3.189 & 3.148 & 3.738 & 3.566 & 3.008 & 2.971 & 3.270 \\
        \racmbasn{}                 & 3.177 & 3.177 & 2.952 & 3.031 & 3.526 & 3.691 & 3.259 \\
    \end{tabular}
    \end{ruledtabular}
\end{table}

\clearpage

\section{Machine-learning force fields (MLFFs)}

\subsection{Force field training}

We trained machine-learning force fields (MLFFs) against total energies, forces and stresses from density functional theory calculations. The training set was automatically constructed using an on-the-fly active learning scheme from dynamical simulations in an \textit{NpT} ensemble~\cite{jinnouchiOntheflyMachineLearning2019, jinnouchiPhaseTransitionsHybrid2019}. Local atomic environments were described using an adaptation of the smooth overlap of atomic positions (SOAP) descriptor~\cite{bartokRepresentingChemicalEnvironments2013}, for which we employed a cutoff for the two-body radial descriptor $\rho_{i}^{\left( 2 \right)}$ of \SI{6.0}{\angstrom}, and a \SI{4.0}{\angstrom} cutoff for the three-body angular descriptor $\rho_{i}^{\left( 3 \right)}$. The atomic positions were broadened using Gaussian distributions with a width of \SI{0.5}{\angstrom}. Both descriptors were expanded on a basis set of spherical Bessel functions and Legendre polynomials, using 8 and 6 Bessel functions for the two-body and three-body descriptors, respectively, with a maximum angular momentum quantum number of $l_{\mathrm{max}} = 2$. The expansion coefficients of this basis set constitute the descriptor for the local atomic environments. To measure the similarity between two local atomic environments, we employed a polynomial kernel function to a power 4, in which the two-body radial and three-body angular descriptor vectors were weighted by 0.1 and 0.9, respectively.

We started MLFF training with a constant temperature simulation at \SI{300}{\K} using the optimized crystal geometry as the starting point. The initial training run was followed by consecutive constant temperature simulations at \SI{100}{\K} and \SI{450}{\K}, each of which used the final positions and velocities as starting point. For all constant temperature simulations the training time was \SI{50}{\ps}. In the final training run we cooled the model systems from \SI{350}{\K} to \SI{50}{\K} over \SI{60}{\ps}, the starting point (i.e. positions and velocities) for these runs was obtained using \SI{10}{\ps} equilibration runs with the respective intermediate force fields. During training the temperature and pressure were controlled using Parinello-Rahman dynamics~\cite{parrinelloCrystalStructurePair1980, parrinelloPolymorphicTransitionsSingle1981} using friction coefficients $\gamma = \SI{5}{\per\ps}$ and $\gamma_{\mathrm{L}} = \SI{5}{\per\ps}$ for the atomic and lattice degrees of freedom, respectively. A time step of $\Delta t = \SI{2}{\fs}$ was used together with an increased mass of the hydrogen atoms of $m_{\ch{H}} = \SI{4}{\amu}$, to increase the sampling rate of new structures. The training sets obtained for the \smbapb{}, \smbasn{} and \smbamixhalf{} MLFFs were put together to form one large training set to train a MLFF for various perovskite compositions of \smbamix{} (0 $\leq$ x $\leq$ 1). To limit the size of the force fields, the number of local reference configuration was limited to 2000 for the individual models and 3000 for the generalized model for arbitrary metal compositions. This limit was enforced by allowing configurations to be discarded once this threshold was reached. Using the training data of the final MLFFs, we refit the models onto faster descriptors without any Bayesian error estimation to speed up the evaluation for large-scale molecular dynamics production runs. The resulting number of training structures and local reference configurations for the different MLFFs can be found in Table~\ref{tab:mlff_details}, the training sets can be found in Ref.~\cite{2dynamicsDatabase2025}.

\begin{table}[htbp]
    \caption{Size of the training set and number of local reference configurations for various machine-learning force fields (MLFFs) for 2D perovskites.}
    \label{tab:mlff_details}
    \begin{ruledtabular}
    \begin{tabular}{ccccccccc}
        \multirow{2}{*}{Perovskite structure}   & \multirow{2}{*}{XC functional} & \multirow{2}{*}{$N_{\mathrm{DFT}}$ (-)}  & \multicolumn{5}{c}{$N_{\mathrm{basis}}$ (-)}                              \\
                                                &                                &                                          & \ch{H}    & \ch{C}    & \ch{N}    & \ch{Sn}   & \ch{I}    & \ch{Pb}       \\ \hline
        \smbapb{}                               & SCAN                           & 972                                      & 2000      & 2000      & 563       & -         & 1240      & 310           \\
        \smbasn{}                               & SCAN                           & 1023                                     & 2000      & 2000      & 552       & 344       & 1279      & -             \\ 
        \smbamixhalf{}                          & SCAN                           & 925                                      & 2000      & 2000      & 557       & 311       & 1558      & 311           \\ \hline
        \smbamix{}                              & SCAN                           & 2920                                     & 3000      & 3000      & 614       & 276       & 1811      & 337           \\ 
    \end{tabular}
    \end{ruledtabular}
\end{table}

\clearpage

\subsection{Force field accuracy}

\begin{figure*}[htbp]
    \includegraphics{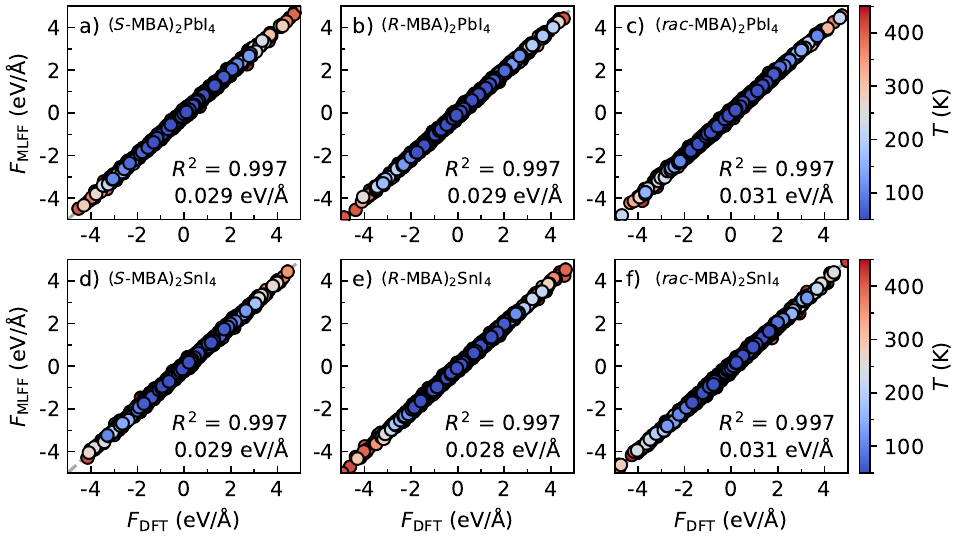}
    \caption{Accuracy of machine-learning force fields (MLFFs) during short heating simulations from \SI{50}{\K} to \SI{450}{\K}. Model performance of MLFF trained on \smbapb{} for (a) \smbapb{}, (b) \rmbapb{} and (c) \racmbapb{}. Model performance of MLFF trained on \smbasn{} for (a) \smbasn{}, (b) \rmbasn{} and (c) \racmbasn{}. In each subfigure the coefficient of determination \rsquared{} and mean absolute error $\mae$ in \SI{}{\eV\per\angstrom} between the forces from the MLFF and DFT are given.}
    \label{fig:mlff_model_accuracy}
\end{figure*}

\begin{figure*}[htbp]
    \includegraphics{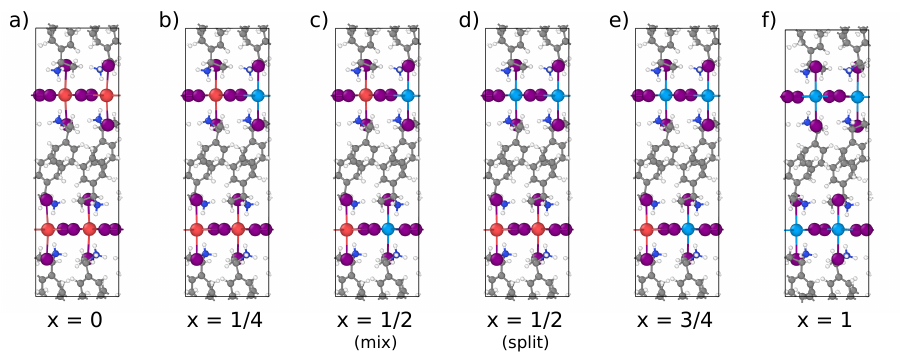}
    \caption{Structures of the various compositions of \smbamix{} used to validate the MLFF model performance against DFT calculations. Structures have a composition of (a) x = 1, (b) x = 1/4, (c) x = 1/2 (mix), (d) x = 1/2 (split), (e) x = 3/4 and (f) x = 1.}
    \label{fig:mlff_model_accuracy_compositions_structures}
\end{figure*}

\begin{figure*}[htbp]
    \includegraphics{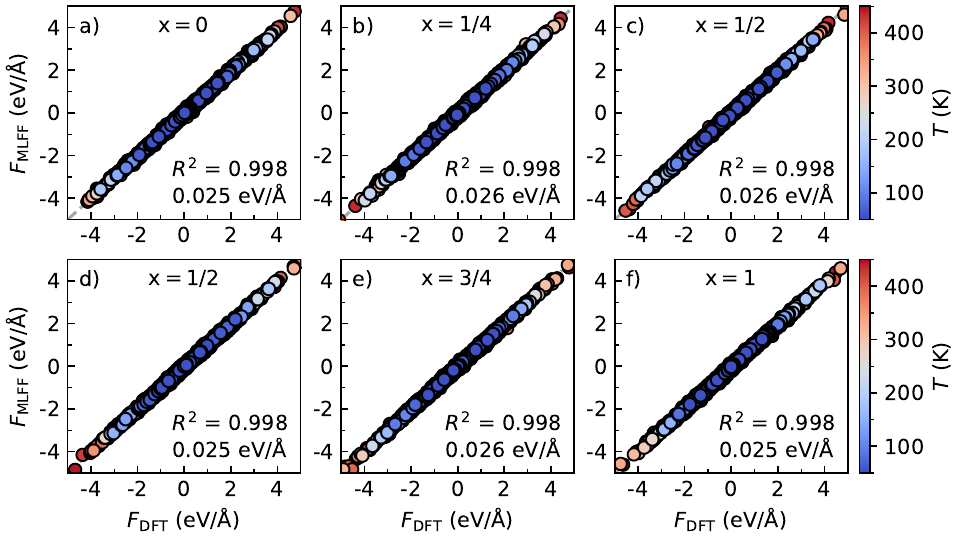}
    \caption{Accuracy of machine-learning force fields (MLFFs) during short heating simulations from \SI{50}{\K} to \SI{450}{\K} for various compositions of \smbamix{}. Model performance of the mixed MLFF for (a) x = 1, (b) x = 1/4, (c) x = 1/2 (mix), (d) x = 1/2 (split), (e) x = 3/4 and (f) x = 1. In each subfigure the coefficient of determination \rsquared{} and mean absolute error $\mae$ in \SI{}{\eV\per\angstrom} between the forces from the MLFF and DFT are given. Structures are shown in Figure~\ref{fig:mlff_model_accuracy_compositions}.}
    \label{fig:mlff_model_accuracy_compositions}
\end{figure*}

To check the ability of the MLFFs to reproduce the forces from DFT calculations, we run short simulations with the final MLFFs obtained from training. During these short simulations, we heated the systems used during training from \SI{50}{\K} to \SI{450}{\K} during \SI{80}{\ps}. We decreased the timestep to $\Delta t$ = \SI{1}{\fs} while decreasing the hydrogen mass to $m_{\ch{H}}$ = \SI{1}{\amu}. The resulting trajectories were divided into 40 equal parts each spanning a temperature range of \SI{10}{\K}. From each part, we randomly selected one frame and evaluated it using both the MLFF and DFT calculations, comparing the force components on all atoms. We note that both the pure models (Figure~\ref{fig:mlff_model_accuracy}), as well as the mixed model (Figure~\ref{fig:mlff_model_accuracy_compositions_structures} and Figure~\ref{fig:mlff_model_accuracy_compositions}) show a high correlation and low error between the MLFF forces and those obtained using DFT.

\clearpage

\section{Mixed metal systems}

\begin{figure*}[htbp]
    \includegraphics{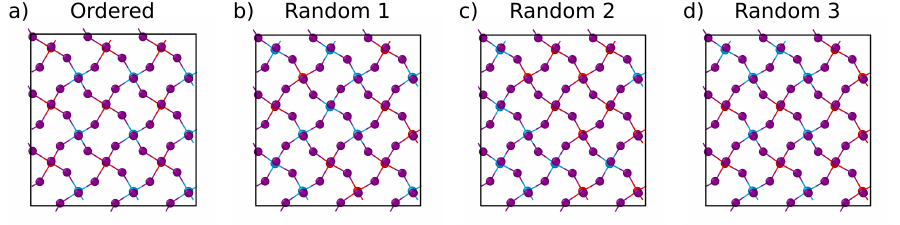}
    \caption{Various structures created through different types of mixing in \smbamixhalf{}. A single inorganic layer is shown for (a) ordered (\singlescrew{}) and (b-d) randomly mixed perovskites (\pone{}). \ch{Pb} and \ch{Sn} are indicated with red and blue spheres, respectively. Due to random mixing, the \ch{Pb}:\ch{Sn} ratio which is 1:1 in the full perovskites, is not necessarily maintained within each inorganic layer.}
    \label{fig:mixed_metal_structures}
\end{figure*}

To check what effect the mixing of the metal cations has on the various properties, such as octahedral distortions ($\sigmatwo$ and $\deltad$) and degree of chirality ($\degreechi$ with $y$ = $\epsA$, $\epsMXpar$, $\epsMXperp$, or $\hbasym$), we compare different \smbamixhalf{} systems. The systems include an ordered mixed and randomly mixed perovskite systems, as shown in Figure~\ref{fig:mixed_metal_structures}. The systems are \supercell{3}{3}{1} supercells, expanded in the lateral directions. All systems have \ch{Pb}:\ch{Sn} in a 1:1 ratio. The ordered mixed structure has \ch{Pb} and \ch{Sn} in a regular checkerboard-like pattern with a \singlescrew{} space group (Figure~\ref{fig:mixed_metal_structures}a), whereas in the randomly mixed structures \ch{Pb}-rich and \ch{Sn}-rich domains decrease the symmetry to a \pone{} space group (Figure~\ref{fig:mixed_metal_structures}b-d).

\clearpage

\section{Degree of chirality}

\begin{figure*}[htbp]
    \includegraphics{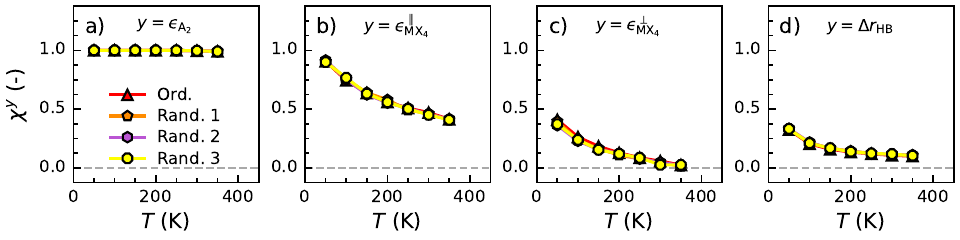}
    \caption{Temperature dependence of degree of chirality in \smbamixhalf{}. The values of $\degreechi$ are shown for (a) $\epsA$, (b) $\epsMXpar$, (c) $\epsMXperp$, and (d) $\hbasym$. The investigated perovskites contain both ordered mixed (Ord.) and randomly mixed (Rand. $i$, with $i$ = 1, 2, or 3) systems.}
    \label{fig:temperature_dependent_degree_of_chirality}
\end{figure*}

To identify the robustness of our findings for mixed metal perovskites, i.e. \smbamixhalf{} ($x$ = 1/2), we compare the temperature dependence of the degree of chirality for mixed perovskites with different metal cation mixing. The resulting comparison, which includes ordered mixed and randomly mixed perovskite systems (Figure~\ref{fig:mixed_metal_structures}), is found in Figure~\ref{fig:temperature_dependent_degree_of_chirality}. We find that $\degreechi$ is not sensitive to the specific metal cation mixing, but to the perovskite composition. All mixed perovskites ($x$ = 1/2), follow a similar temperature dependence for all descriptors, i.e. $\epsA$, $\epsMXpar$, $\epsMXperp$, and $\hbasym$, that falls between that seen in \smbapb{} ($x$ = 0) and \smbasn{} ($x$ = 1).

\clearpage

\section{Head group reorientations}

As a probe for the chirality transfer occurring between the organic cations and the inorganic framework, we study the dynamics of the \ch{NH3^{+}} head groups in more detail. We do this through the autocorrelation function of the $\ch{N}-\ch{H}$ bonds, which is defined as
\begin{equation}
    \autocorrelationNH \left( t \right) = \left \langle \frac{1}{N} \sum^{N}_{i = 1} \mathbf{\hat{u}}_{i} \left( t_{0} \right) \cdot \mathbf{\hat{u}}_{i} \left( t_{0} + t \right) \right \rangle_{t_{0}},
\end{equation}
where $\mathbf{\hat{u}}_{i} = \mathbf{r}_{i} / | \mathbf{r}_{i} |$ is the normalized orientation vector of the $i$\textsuperscript{th} $\ch{N}-\ch{H}$ bond, $N$ the number of bond vectors, $t$ the time delay and $t_{0}$ the time origin. The autocorrelation function is averaged over different time origins, as indicated with $\langle \cdots \rangle_{t_{0}}$. The time at which autocorrelation function falls of to a value of $\autocorrelationNH \left( \tauNH \right) = e^{-1}$ is defined as the orientation lifetime $\tauNH$.

\begin{figure*}[htbp]
    \includegraphics{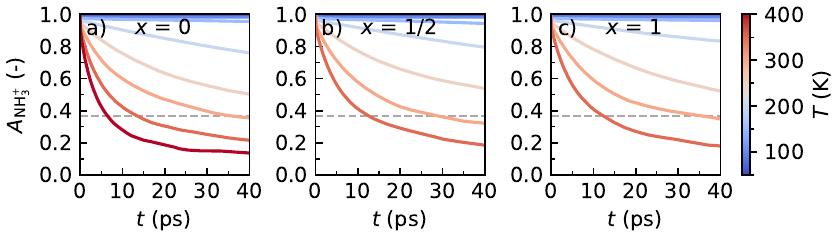}
    \caption{Cation head group orientational autocorrelation function $\autocorrelationNH$ for various compositions of \smbamix{}, with (a) $x$ = 0, (b) $x$ = 1/2, and (c) $x$ = 1.}
    \label{fig:headgroup_composition}
\end{figure*}

\begin{figure*}[htbp]
    \includegraphics{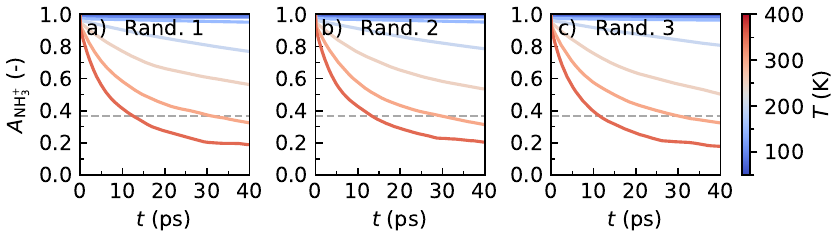}
    \caption{Cation head group orientational autocorrelation function $\autocorrelationNH$ for (a-c) various randomly mixed \smbamixhalf{} perovskites.}
    \label{fig:headgroup_random}
\end{figure*}

By identifying the orientational lifetimes of the various perovskite compositions, we can characterize the persistence of hydrogen bonds in these compounds. Table~\ref{tab:autocorrelation_times} provides an overview of the values of $\tauNH$ at \SI{350}{\K}, which correspond to the autocorrelation functions shown in Figure~\ref{fig:headgroup_composition} and Figure~\ref{fig:headgroup_random}. Irrespective of the composition and if relevant type of metal cation mixing, all compounds show a similar persistence in the hydrogen bonds, indicating the strength of the hydrogen bonds is the same across the compounds.

\begin{table}[htbp]
    \caption{Orientational lifetimes of $\ch{N}-\ch{H}$ bonds in \smbamix{} perovskites at \SI{350}{\K}.}
    \label{tab:autocorrelation_times}
    \begin{ruledtabular}
    \begin{tabular}{cccc}
        Composition                 & $\tauNH$ (\SI{}{\ps})         & Composition                 & $\tauNH$ (\SI{}{\ps})         \\ \hline
        $x$ = 0                     & 13.5                          & $x$ = 1/2 (Rand. 1)         & 12.6                          \\
        $x$ = 1/2 (Ord.)            & 12.5                          & $x$ = 1/2 (Rand. 2)         & 13.3                          \\ 
        $x$ = 1                     & 12.4                          & $x$ = 1/2 (Rand. 3)         & 11.4                          \\
    \end{tabular}
    \end{ruledtabular}
\end{table}

\clearpage

\section{Phonon calculations}

\subsection{Density of states (DOS)}

\begin{figure*}[htbp]
    \includegraphics{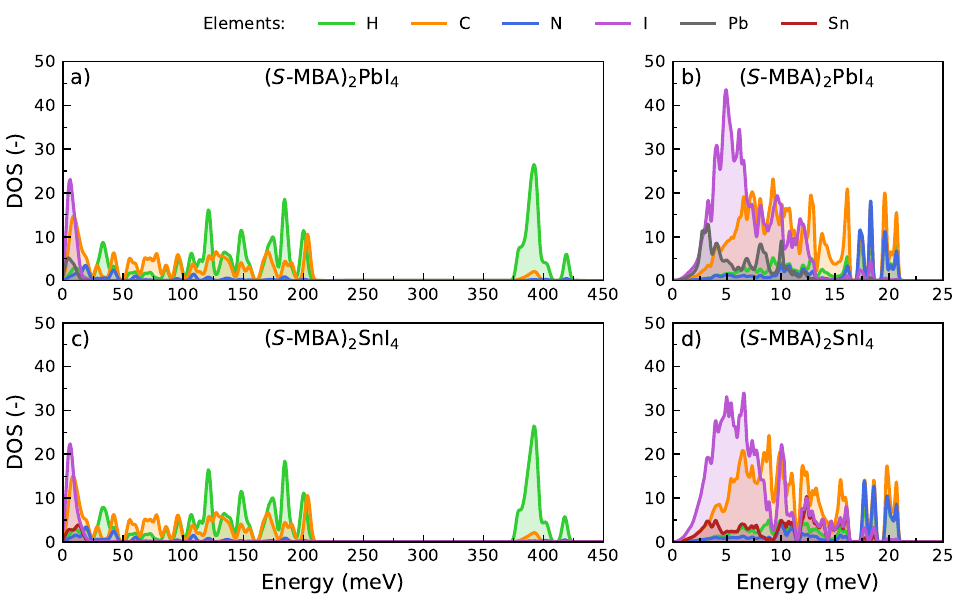}
    \caption{Phonon density of states (DOS). The (a) full DOS and (b) zoomed in DOS of the lower energy region (0 - \SI{25}{\meV}) of \smbapb{}. The (c) full DOS and (d) zoomed in DOS of the lower energy region (0 - \SI{25}{\meV}) of \smbasn{}. A Gaussian broadening of \SI{0.1}{\meV} and \SI{2.0}{\meV} were used in the full and zoomed in DOS, respectively.}
    \label{fig:phonon_dos}
\end{figure*}

The density of states of both perovskites, shown in Figure~\ref{fig:phonon_dos}, exhibit similar characteristics, featuring three regions: (i) low energy region (0 - \SI{25}{\meV}), (ii) intermediate energy region (25 - \SI{210}{\meV}), and (iii) a high energy region (375 - \SI{425}{\meV}). In the low energy region (i), primarily the inorganic framework vibrates, with some translational and rotational contributions from the organic cations. The intermediate (ii) and high energy region (iii) are the result of motion of the organic cations. Despite the many similarities in the overall phonon DOS (Figure~\ref{fig:phonon_dos}a and Figure~\ref{fig:phonon_dos}c), the fine structure of the low energy region does exhibit differences (Figure~\ref{fig:phonon_dos}b and Figure~\ref{fig:phonon_dos}d), in particular the vibrations of the halide (\ch{I}) and metal (\ch{Pb} and \ch{Sn}) ions, which can be attributed to the differences in the octahedral geometries between the two perovskites.

\clearpage

\subsection{Group velocities}

From the acoustic phonons, we calculate the phonon group velocity, which is defined as
\begin{equation}
    \label{eq:group_velocity}
    \vg \left( \mathbf{q}, \sigma \right) = \nabla_{\mathbf{q}} \omega_{\mathbf{q}, \sigma},
\end{equation}
for which we take the gradient of the angular frequency $\omega_{\mathbf{q}, \sigma}$ as a function of the wavevector $\mathbf{q}$. This is done arbitrarily close to the \textGamma{}-point along the high-symmetry paths. The group velocities in the in-plane (\textGamma{}$-$X and \textGamma{}$-$Y paths) and out-of-plane (\textGamma{}$-$Z path) directions. The group velocities and their average in each direction are reported for \smbapb{} and \smbasn{} in Table~\ref{tab:acoustic_group_velocities}. Both chiral 2D perovskites exhibit a low anisotropy in the group velocity of the phonons, as seen in earlier work~\cite{liRemarkablyWeakAnisotropy2021}.

\begin{table}[htbp]
    \caption{Group velocities of the acoustic phonons in various 2D perovskites.}
    \label{tab:acoustic_group_velocities}
    \begin{ruledtabular}
    \begin{tabular}{cccccc}
        \multirow{2}{*}{Perovskite}             & \multirow{2}{*}{Mode} & \multicolumn{3}{c}{Group velocities}                                                                      \\
                                                &                       & $\vgroup{X}$  (\SI{}{\m\per\s})   & $\vgroup{Y}$  (\SI{}{\m\per\s})   & $\vgroup{Z}$  (\SI{}{\m\per\s})   \\ \hline
        \multirow{4}{*}{\smbapb{}}              & 1                     & 1173.0                            & 1169.9                            & 1170.8                            \\
                                                & 2                     & 1557.8                            & 1558.4                            & 1174.9                            \\ 
                                                & 3                     & 2867.1                            & 2764.9                            & 2749.9                            \\ 
                                                & Average               & 1866.0                            & 1831.1                            & 1698.5                            \\ \hline
        \multirow{4}{*}{\smbasn{}}              & 1                     & 1113.3                            & 1105.3                            & 1105.8                            \\
                                                & 2                     & 1136.9                            & 1138.0                            & 1110.3                            \\
                                                & 3                     & 2659.8                            & 2518.8                            & 2921.7                            \\ 
                                                & Average               & 1636.7                            & 1587.4                            & 1712.6                            \\
    \end{tabular}
    \end{ruledtabular}
\end{table}

Interestingly, comparing the different compositions with each other, we find that \smbapb{} has considerably larger in-plane phonon group velocities than \smbasn{} has. We hypothesize that this discrepancy is caused by the highly anisotropic bonding in the in-plane direction of the metal halide octahedra in the \ch{Sn}-based perovskite (Table~\ref{tab:bond_lengths}). The long equatorial bonds, i.e. \SI{3.74}{\angstrom} and \SI{3.55}{\angstrom}, provide a weak coupling between atoms in the inorganic layers, substantially reducing the in-plane propagation of phonons and thus lower group velocities. Interestingly, this coupling is weak enough to negate the increased group velocity resulting from the smaller mass of \ch{Sn}.

\clearpage

\subsection{Effect of mixing on phonons}

\begin{figure*}[htbp]
    \includegraphics{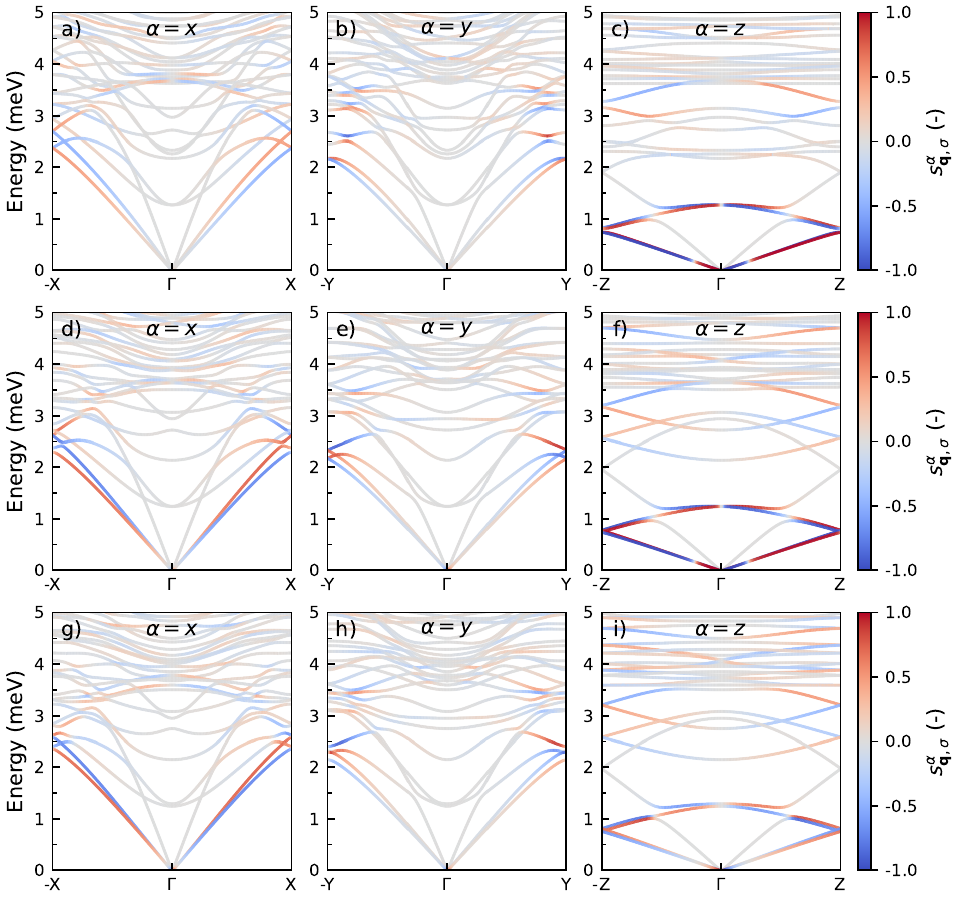}
    \caption{Circularly polarized phonon spectra of \smbamixhalf{} for different metal configurations. Phonon spectra for structure with \screw{}-axis along the (a-c) $x$-axis, (d-f) $y$-axis, and (g-i) $z$-axis.}
    \label{fig:phonon_spectra_mixing}
\end{figure*}

To examine how metal cation mixing influences phonon chirality, we present the phonon dispersion for varying mixing ratios in the unit cell of \smbamixhalf{} (Figure~\ref{fig:phonon_spectra_mixing}). The unit cell allows for three distinct metal cation configurations, all adopting the \singlescrew{} space group with the \screw{}-axis aligned along the Cartesian coordinates, i.e. $x$, $y$, and $z$. The resulting phonon dispersions and their associated chirality resemble a mix of those observed in the pure compounds, with subtle variations introduced by the specific metal configuration.

\clearpage

\subsection{Model sensitivity}

\begin{figure*}[htbp]
    \includegraphics{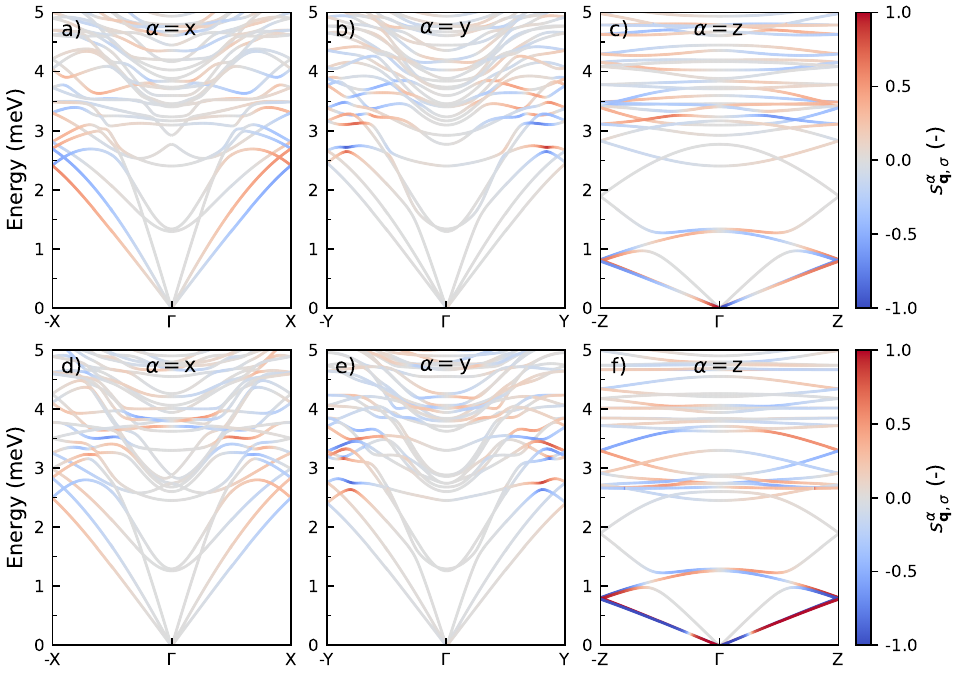}
    \caption{Circularly polarized phonon spectra of \smbapb{}. Phonon spectra for (a-c) the pure model trained on only \smbapb{} data and (d-f) the mixed model trained on \smbamix{} ($x$ = 0, 1/2, and 1) data.}
    \label{fig:phonon_spectra_comparison}
\end{figure*}

Two MLFFs were trained for \smbapb{}, one trained on only \smbapb{} data and one trained against data of \smbamix{} ($x$ = 0, 1/2, and 1). By comparing the circularly polarized phonon dispersions of these two models, we can check their sensitivity to the MLFFs. The phonon spectra (Figure~\ref{fig:phonon_spectra_comparison}) exhibit some differences in the frequencies of the phonon branches as well as the magnitude of their circular polarization. Nevertheless, both models result in phonon spectra with an overall similar character.

\clearpage

\section{Angular momentum generation}

\subsection{Experimental observation}

To evaluate whether the angular momentum generated under a temperature gradient is large enough to be experimentally observable, we calculate the rigid-body rotation induced in a macroscopic crystal. For this purpose, we use the method described by \textcite{hamadaPhononAngularMomentum2018}. From the conservation of angular momentum, it follows that the total angular momentum generated ($\mathbf{J}^{\mathrm{ph}}$) and the angular momentum from the rotation of a rigid-body ($\mathbf{J}^{\mathrm{rb}}$) sum to zero, for a cubic sample of size $V = L^{3}$, as $\mathbf{J}^{\mathrm{ph}} L^{3} + \mathbf{J}^{\mathrm{rb}} L^{3} = 0$. For a temperature gradient along the $x$-direction this yields
\begin{equation}
    \label{eq:rigid-body_x_direction}
    J^{\mathrm{rb},x} L^{3} = I \omega = - \alpha^{xx} \frac{\Delta T}{L} L^{3}
\end{equation}
which, using the inertial moment for a cube of mass $M = \rho V$, $I = M L^{2} / 6$, provides us with an expression for the angular velocity of
\begin{equation}
    \label{eq:angular_velocity}
    \omega = - \frac{6 \alpha^{xx}}{\rho} \frac{\Delta T}{L^{3}},
\end{equation}
where $\rho$ is the density of the material. The calculated values for \smbapb{} and \smbasn{} are shown in Table~\ref{tab:rigid-body_rotation} for cubic samples with sides of length $L = \SI{1}{\micro\meter}$ and $L = \SI{100}{\micro\meter}$, using $\tau = \SI{1}{\ps}~\cite{acharyyaIntrinsicallyUltralowThermal2020}$ as a conservative estimate. We note that the resulting values for \smbapb{} are in the same range as earlier values for \ch{GaN}, \ch{Te}, and \ch{Se}~\cite{hamadaPhononAngularMomentum2018}, and the effect should therefore be large enough to be experimentally observable.

\begin{table}[htbp]
    \caption{Rigid-body rotation for chiral 2D perovskites under application of a temperature gradient along the $x$-direction. Values are computed for cubic samples with sides $L = \SI{1}{\micro\meter}$ and $L = \SI{100}{\micro\meter}$.}
    \label{tab:rigid-body_rotation}
    \begin{ruledtabular}
    \begin{tabular}{lcc}
        Quantity                                                                                & \smbapb{}                         & \smbasn{}                         \\ \hline
        $\rho$ (\SI{}{\kg\per\m\cubed})                                                         & \SI{2.72E3}{}                     & \SI{2.44E3}{}                     \\
        $\alpha^{xx} / \tau$ (\SI{}{\joule\per\m\squared\per\kelvin}) (\@$T = \SI{300}{\K}$)    & \SI{8.17E-08}{}                   & \SI{5.90E-10}{}                   \\
        $\tau$ (\SI{}{\ps})                                                                     & 1                                 & 1                                 \\
        $\Delta T$ (\SI{}{\kelvin})                                                             & 10                                & 10                                \\
        $\omega$ (\SI{}{\per\s}) ($L = \SI{1}{\micro\meter}$)                                   & -\SI{1.8E-3}{}                    & -\SI{1.5E-5}{}                    \\
        $\omega$ (\SI{}{\per\s}) ($L = \SI{100}{\micro\meter}$)                                 & -\SI{1.8E-9}{}                    & -\SI{1.5E-11}{}                   \\
    \end{tabular}
    \end{ruledtabular}
\end{table}

\clearpage

\subsection{Model sensitivity}

\begin{figure*}[htbp]
    \includegraphics{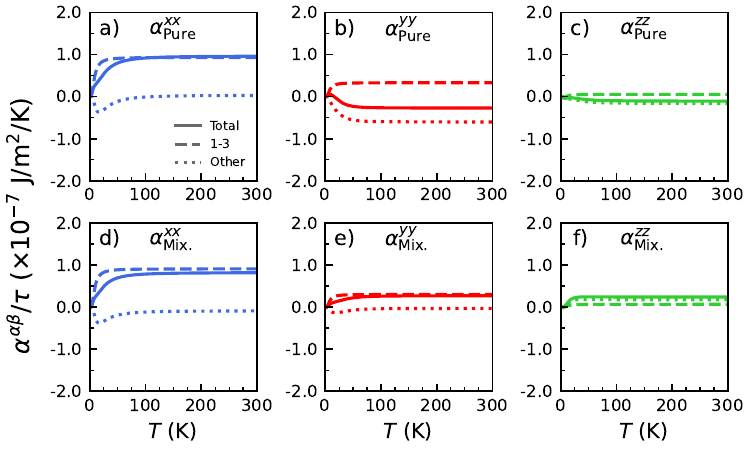}
    \caption{Temperature dependence of the (a) $\alpha^{xx}$, (b) $\alpha^{yy}$, and (c) $\alpha^{zz}$ response tensors for \smbapb{} with the pure model. Temperature dependence of the (d) $\alpha^{xx}$, (e) $\alpha^{yy}$, and (f) $\alpha^{zz}$ response tensors for \smbapb{} with the mixed model.}
    \label{fig:phonon_pam_model_comparison}
\end{figure*}

Furthermore, we also assessed the sensitivity of the generated angular momentum to the different MLFFs. Using both the pure \smbapb{} model and the mixed \smbamix{} model, we computed the response tensor $\alpha^{\alpha \beta}$ for \smbapb{}. The resulting response tensors are shown in Figure~\ref{fig:phonon_pam_model_comparison}. Despite variations in the phonon spectra (Figure~\ref{fig:phonon_spectra_comparison}), both models result in a similar response tensor. The large component ($\alpha^{xx}$) is large and of similar magnitude for both models, and the small components ($\alpha^{yy}$ and $\alpha^{zz}$) show more variations but remain small for both.

\clearpage

\bibliography{si}